\documentclass[12pt,english]{article}
\usepackage{graphicx,array}
\usepackage{cite}
\usepackage{color}
\usepackage{latexsym}
\usepackage{amsthm}
\usepackage{amsmath}
\usepackage[titletoc]{appendix}
\usepackage{enumitem}
\usepackage{amssymb}
\usepackage{hyperref}
\usepackage[hang,flushmargin]{footmisc} 

\def\simgt{\mathrel{\lower2.5pt\vbox{\lineskip=0pt\baselineskip=0pt
           \hbox{$>$}\hbox{$\sim$}}}}
\def\simlt{\mathrel{\lower2.5pt\vbox{\lineskip=0pt\baselineskip=0pt
           \hbox{$<$}\hbox{$\sim$}}}}

\newcommand{\be}{\begin{equation}}
\newcommand{\ee}{\end{equation}}
\newcommand{\bea}{\begin{eqnarray}}
\newcommand{\eea}{\end{eqnarray}}
\newcommand{\Eq}[1]{Eq.~(\ref{#1})}
\newcommand{\Eqs}[2]{Eqs.~(\ref{#1}) and (\ref{#2})}
\newcommand{\Sec}[1]{Sec.~\ref{#1}}

\newcommand{\Fig}[1]{Fig.~\ref{#1}}

\newcommand{\Ref}[1]{Ref.~\cite{#1}}

\newcommand{\mPl}{m_{\rm Pl}}

\newcommand*\oline[1]{%
  \vbox{%
    \hrule height 0.5pt
    \kern0.68ex
    \hbox{%
      \kern-0.1em
      \ifmmode#1\else\ensuremath{#1}\fi
      \kern-0.1em
    }
  }
}

\definecolor{nicered}{rgb}{0.7,0.1,0.1}
\definecolor{nicegreen}{rgb}{0.1,0.5,0.1}
\hypersetup{
 colorlinks,citecolor=black,linkcolor=black,urlcolor=black}

\begin{document}
\interfootnotelinepenalty=10000
\baselineskip=18pt
\hfill CALT-TH-2015-062
\hfill

\vspace{1.5cm}
\thispagestyle{empty}
\begin{center}
{\LARGE\bf
Positive Signs in Massive Gravity
}\\
\bigskip\vspace{1.5cm}{
{\large Clifford Cheung and Grant N. Remmen}
} \\[7mm]
 {\it Walter Burke Institute for Theoretical Physics, \\[-1mm]
    California Institute of Technology, Pasadena, CA 91125}\let\thefootnote\relax\footnote{e-mail: \url{clifford.cheung@caltech.edu}, \url{gremmen@theory.caltech.edu}} \\
 \end{center}
\bigskip
\centerline{\large\bf Abstract}

\begin{quote} \small
We derive new constraints on massive gravity from unitarity and analyticity of scattering amplitudes.  Our results apply to a general effective theory defined by Einstein gravity plus the leading soft diffeomorphism-breaking corrections.  We calculate scattering amplitudes for all combinations of tensor, vector, and scalar polarizations.  The high-energy behavior of these amplitudes prescribes a specific choice of couplings that ameliorates the ultraviolet cutoff, in agreement with existing literature.  We then derive consistency conditions from analytic dispersion relations, which dictate positivity of certain combinations of parameters appearing in the forward scattering amplitudes.  These constraints exclude all but a small island in the parameter space of ghost-free massive gravity.
While the theory of the ``Galileon'' scalar mode alone is known to be inconsistent with positivity constraints, this is remedied in the full massive gravity theory.
\end{quote}

\setcounter{footnote}{0}

\newpage
\tableofcontents
\newpage

\section{Introduction}

Local symmetry breaking is a central concept in quantum field theory with a rich theoretical structure and ubiquitious applications to natural phenomena.  While this subject is textbook material in the context of gauge theories, its gravitational analogue remains an active field of study.  In particular, theories of massive gravity have spawned an extensive body of literature analyzing its formal aspects and phenomenology (see \Ref{Kurt} and references therein).  

In this paper, we present new constraints on the parameter space of massive gravity coming from the consistency of scattering amplitudes. For the sake of generality, we assume an effective theory for massive gravity comprised of general relativity plus soft diffeomorphism-breaking corrections proportional to the graviton mass \cite{Kurt}.   The theory contains five degrees of freedom: two tensors, two vectors, and one scalar, which is known in the literature as the ``Galileon".  Importantly, we work in unitarity gauge so that the tensor, vector, and scalar modes are manipulated together as a multiplet rather than as decoupled states in the limit of Goldstone equivalence~\cite{Nima}.  

To eliminate ghost modes, we restrict to the parameter space of ghost-free massive gravity \cite{dRG,dRGT}, which is the non-linear generalization of the Fierz-Pauli tuning for the graviton mass.  Notably, ghost-free massive gravity has a parametrically higher cutoff than a generic massive gravity theory~\cite{dRG} and the resulting action has two free coupling constants, $(c_3, d_5)$ \cite{Kurt}.

After an intensive computation, we arrive at lengthy expressions for the general tree-level amplitude for the scattering of massive gravitons. As we will show in detail, analyticity and unitarity place positivity constraints on the coefficients that appear in the forward amplitude. Imposing positivity on {\it all possible} graviton scattering processes, we sculpt an allowed region in $(c_3,d_5)$. For external states that are described by pure tensor, vector, or scalar polarizations---which we dub ``definite-helicity'' states---we obtain the excluded colored regions shown in \Fig{fig:exclusions}. Expanding to the scattering of arbitrary superpositions of tensors, vectors, and scalars---which we dub ``indefinite-helicity'' states---we derive more stringent constraints, leaving a compact allowed region in $(c_3, d_5)$ permitted by unitarity and analyticity shown in \Fig{fig:dynamical}.  

While this result excludes much of the parameter space of massive gravity, it is actually a boon to the Galileon, which as a stand-alone effective theory actually fails analyticity bounds \cite{Rattazzi,Positivity,IRUV}.  However, since this failure is marginal, corrections to the limit of Goldstone equivalence can tip the balance to restore analyticity in the theory.  Thus, non-analyticity of the original Galileon may be corrected by embedding it into the full theory of massive gravity.

The structure of this paper is as follows.  In \Sec{sec:theory}, we describe a general effective theory for massive gravity.  Next, we compute the massive graviton scattering amplitudes in \Sec{sec:amplitudes} and verify that they are consistent with existing literature. Finally, in \Sec{sec:constraints} we present our new bounds from analytic dispersion relations, discuss implications in \Sec{sec:galileon}, and conclude in \Sec{sec:conclusions}.

\section{Effective Theory for Massive Gravity}

\label{sec:theory}

We consider a general effective theory for massive gravity defined by the Einstein-Hilbert term plus soft diffeomorphism-breaking operators \cite{Kurt}.  This starting point is familiar from other contexts, {\it e.g.}, soft breaking of gauge symmetry or supersymmetry.  In such instances, hard symmetry breaking should be avoided since it is radiatively unstable.  The action for the massive gravity effective theory is
\be 
S = \frac{\mPl^2}{2} \int {\rm d}^4 x  \sqrt{-g} \left[ R - \frac{m^2}{4} V(g,h) \right].\label{eq:action}
\ee
The metric is $g_{\mu\nu} = \eta_{\mu\nu} + h_{\mu\nu} $, where $\eta_{\mu\nu}$ is the flat metric in mostly $+$ signature and $h_{\mu\nu}$ corresponds to the graviton. Here $m$ is the soft breaking parameter, to be identified with the graviton mass shortly.  Throughout, $\mPl=1/\sqrt{8\pi G}$ is the reduced Planck mass.

The graviton potential terms take the general form
\be 
\begin{aligned}
V(g,h) =&\; V_2(g,h)+V_3(g,h)+V_4(g,h)+ \cdots \\
V_2(g,h) =&\; +b_1\langle h^2 \rangle +b_2\langle h \rangle^2  \\
V_3(g,h) =&\; +c_1\langle h^3 \rangle +c_2 \langle h \rangle^2 \langle h\rangle + c_3 \langle h\rangle^3 \\
V_4(g,h) =&\; +d_1\langle h^4 \rangle +d_2 \langle h^3 \rangle \langle h\rangle +d_3 \langle h^2 \rangle^2 +d_4 \langle h^2 \rangle \langle h\rangle^2 + d_5 \langle h\rangle^4,
\label{eq:Vdef}
\end{aligned}
\ee
where angle brackets denote full metric contractions: $\langle h \rangle = g^{\mu\nu} h_{\mu\nu}$, $\langle h^2 \rangle = g^{\mu\nu} h_{\nu \rho} g^{\rho \sigma} h_{\sigma \mu}$, etc.

We assume the Fierz-Pauli form for the graviton mass terms,
\be 
b_1  = -b_2 =1,\label{eq:FPtuning}
\ee
so the linearized theory describes a massive graviton with five polarizations: two tensors, two vectors, and one scalar.  Without the Fierz-Pauli tuning in \Eq{eq:FPtuning}, the Hamiltonian loses a constraint, activating a scalar ghost degree of freedom \cite{Kurt}.

At the non-linear level, however, numerous pathologies arise.  For example, Boulware and Deser \cite{BDghost} observed that a dangerous ghost degree of freedom is reintroduced in non-trivial backgrounds.  Moreover, the high-energy behavior of the amplitude signals a parametrically low cutoff $\Lambda_5$ \cite{Nima}, where for later convenience we define
\be 
\Lambda_n = (m^{n-1} \mPl)^{1/n}.
\ee
More recently, it was observed that the Boulware-Deser ghost can be eliminated with the proper choice of parameters \cite{dRG,dRGT,Hassan:2011hr}.  In particular, working in the high-energy theory of scalars, the couplings at each power in the graviton can be chosen to yield total derivative interactions.  For example, in \Eq{eq:Vdef} this parameter choice corresponds to
\be 
\begin{aligned}
c_1 &= 2c_3 + \frac{1}{2},   &\quad 
c_2 &= -3 c_3 -\frac{1}{2}, \\
d_1 &= -6 d_5 + \frac{3}{2} c_3 + \frac{5}{16}, &
d_2 &= 8d_5 -\frac{3}{2}c_3 -\frac{1}{4}, \\
d_3 &= 3d_5 - \frac{3}{4} c_3 - \frac{1}{16},  & 
d_4 &= -6d_5 +\frac{3}{4} c_3,
\label{eq:params}
\end{aligned}
\ee
with $c_3$ and $d_5$ free parameters.  The resulting theory is a non-linear generalization of the Fierz-Pauli term.  Moreover, the theory enjoys a parametrically higher cutoff $\Lambda_3$ \cite{dRG,Nima},
since the parameter choice eliminates dangerous scalar self-interactions.

\section{Calculation of Scattering Amplitudes}

\label{sec:amplitudes}

For our analysis, we have computed the general tree-level amplitude for massive graviton scattering.  In what follows, we describe the setup and notation of our amplitudes calculation, followed by a set of consistency checks for our final expressions.

\subsection{Setup and Notation} \label{sec:Setup}

A massive graviton has a momentum vector $k_\mu$ satisfying $k_\mu k^\mu = -m^2$.  To construct a basis of polarization tensors, we decompose the space orthogonal to $k_\mu$ in terms of a basis of three polarization vectors $\epsilon^i_\mu$ satisfying
\be 
k^\mu \epsilon_\mu^i   =0
\ee 
and split according to transverse ($i=1,2$) and longitudinal ($i=3$) polarizations.  For example, in a frame in which $k_\mu = (\omega,0,0,k) $ and  $\omega = \sqrt{k^2 +m^2}$, the polarization vectors satisfy
\be 
\begin{aligned}
\epsilon_{\mu}^1 &= (0,1,0,0)\\
\epsilon_{\mu}^2 &= (0,0,1,0)\\
\epsilon_{\mu}^3 &= \frac{1}{m}(k,0,0,\omega), \label{eq:polexplicitbasis}
\end{aligned}
\ee
with the normalization $\epsilon^i_\mu \epsilon^{j\mu} = \delta^{ij}$.
By construction, at high energies $\epsilon^3_\mu \sim k_\mu /m$, which is the Goldstone equivalence limit.

Next, we construct a basis of five polarization tensors $\epsilon^i_{\mu\nu}$, which are symmetric and satisfy the transverse traceless conditions
\be 
k^\mu \epsilon_{\mu\nu}^i = \epsilon_\mu^{i\;\,\mu} =0,
\ee 
normalized to $\epsilon^i_{\mu\nu} \epsilon^{j\mu\nu} = \delta^{ij}$.    
Here the tensor ($i=1,2$), vector ($i=3,4$), and scalar ($i=5$) polarizations are\footnote{The overall phase of each polarization is unphysical, but we include a factor of $i$ in the vector polarizations to manifest their odd parity under charge conjugation. }
\be 
\begin{aligned}
\epsilon^1_{\mu\nu} &=  \frac{1}{\sqrt{2}} (\epsilon^1_\mu  \epsilon^1_\nu -\epsilon^2_\mu  \epsilon^2_\nu),& \qquad \epsilon^2_{\mu\nu}  &= \frac{1}{\sqrt{2}} (\epsilon^1_\mu  \epsilon^2_\nu +\epsilon^2_\mu  \epsilon^1_\nu),\\
\epsilon^3_{\mu\nu} &=  \frac{i}{\sqrt{2}} (\epsilon^1_\mu  \epsilon^3_\nu +\epsilon^3_\mu  \epsilon^1_\nu),& \qquad  \epsilon^4_{\mu\nu} &= \frac{i}{\sqrt{2}}(\epsilon^2_\mu  \epsilon^3_\nu +\epsilon^3_\mu  \epsilon^2_\nu), \\
 \epsilon^5_{\mu\nu} &= \sqrt{\frac{3}{2}}\left(\epsilon^3_\mu \epsilon^3_\nu - \frac{1}{3}\Pi_{\mu\nu}\right), && \label{eq:allpols}
\end{aligned}
\ee
where we have defined the projection operator 
\be 
\Pi_{\mu\nu} = \eta_{\mu\nu} + \frac{k_\mu k_\nu}{m^2}. \label{eq:proj}
\ee
The polarizations satisfy the completeness relation,
\be
\sum_i \epsilon^i_{\mu\nu} \epsilon^{i*}_{\rho\sigma} = \frac{1}{2}\left(\Pi_{\mu\rho}\Pi_{\nu\sigma} + \Pi_{\mu\sigma}\Pi_{\nu\rho}\right)-\frac{1}{3}\Pi_{\mu\nu}\Pi_{\rho\sigma},
\ee
where the right side is the massive graviton propagator numerator. We will often denote the tensor, vector, and scalar polarizations schematically as $T$, $V$, and $S$, respectively.  The last is also known in the literature as the Galileon \cite{DGP,Rattazzi,Positivity,Luty:2003vm}.

In terms of the explicit frame used in \Eq{eq:polexplicitbasis}, the polarization tensors are
\be 
\begin{aligned}
\epsilon^{1}_{\mu\nu} &= \frac{1}{\sqrt{2}} \begin{pmatrix} 0 & 0 & 0 & 0 \\ 0 & 1 & 0 & 0\\ 0 & 0 & -1 & 0 \\ 0 & 0 & 0 & 0 \end{pmatrix},& \qquad \epsilon^{2}_{\mu\nu} &=   \frac{1}{\sqrt{2}} \begin{pmatrix} 0 & 0 & 0 & 0 \\ 0 & 0 & 1 & 0\\ 0 & 1 & 0 & 0 \\ 0 & 0 & 0 & 0 \end{pmatrix},\\
\epsilon^{3}_{\mu\nu} &= \frac{i}{\sqrt{2}m} \begin{pmatrix} 0 & k & 0 & 0 \\ k & 0 & 0 &\omega\\ 0 & 0 & 0 & 0 \\ 0 & \omega & 0 & 0 \end{pmatrix} ,& \qquad \epsilon^{4}_{\mu\nu} &= \frac{i}{\sqrt{2}m} \begin{pmatrix} 0 & 0 & k & 0 \\ 0 & 0 & 0 &0\\ k & 0 & 0 & \omega \\ 0 & 0 & \omega & 0 \end{pmatrix}, \\
\epsilon^5_{\mu\nu} &= \sqrt{\frac{2}{3}}\frac{1}{m^2} \begin{pmatrix} k^2 & 0 & 0 & k \omega \\ 0 & -m^2/2 & 0 & 0 \\0 & 0 & -m^2/2 & 0 \\  k \omega & 0 & 0 & \omega^2 \end{pmatrix},\hspace*{-1in} & &
\end{aligned}
\ee
which can come in handy for explicit calculations.

The general scattering amplitude of massive gravitons, $M(ABCD)$, depends on the Mandelstam invariants $(s,t)$ together with four external polarization tensors,
\be 
\begin{aligned}
\epsilon^A_{\mu\nu} &= \sum_i \alpha_i \epsilon^i_{\mu\nu} ,& \qquad \epsilon^B_{\mu\nu} &= \sum_i\beta_i \epsilon^i_{\mu\nu}, \\
\epsilon^C_{\mu\nu} &= \sum_i\gamma_i \epsilon^i_{\mu\nu},& \qquad  \epsilon^D_{\mu\nu} &= \sum_i\delta_i \epsilon^i_{\mu\nu} ,
\end{aligned}
\ee
where $\alpha, \beta, \gamma, \delta$ are unit vectors.

To determine constraints, we restrict to forward, crossing-symmetric amplitudes.  The forward limit implies $t= 0$, which is a regular kinematic regime, as the graviton mass regulates all infrared singularities.  Meanwhile, the constraint of crossing symmetry requires that
\bea
\epsilon^{C*}_{\mu\nu} = \epsilon^A_{\mu\nu}  \quad {\rm and} \quad \epsilon^{D*}_{\mu\nu} = \epsilon^{B}_{\mu\nu}.
\eea
Thus, the general scattering amplitude is a function of $(s, t,\alpha,\beta,\gamma,\delta)$ while the forward, crossing-symmetric amplitude is a function of $(s,\alpha,\beta)$. In order to maintain crossing symmetry simultaneously with the forward limit, we must assume linear polarizations for the external states \cite{us}, which means that the vectors $\alpha$ and $\beta$ are real. 

We have calculated the massive graviton scattering amplitude at general kinematics using the above definitions of the external polarization tensors, together with the Feynman rules extracted from \Eq{eq:Vdef} after going to canonical normalization where $h_{\mu\nu}$ is rescaled by $\mPl/2$.   As our amplitudes expressions are prohibitively long, we include them as supplemental material.

\subsection{Consistency Checks}

To verify consistency we have studied the high-energy behavior for ``definite-helicity'' gravitons, which are strictly $T$, $V$, or $S$.   From power counting, we know that the massive graviton modes enter the action as $T \sim \partial V \sim \partial\partial S$, so the high-energy behavior of amplitudes at fixed angle is 
\be
\begin{aligned}
&M(TTTT) \sim s,&  &M(TVTV) \sim s^2,& &M(TSTS) \sim s^3 , \\
&M(VVVV) \sim s^3,&  &M(VSVS) \sim s^4,& &M(SSSS) \sim s^5 .
\end{aligned}
\ee
Our explicit amplitude expressions agree with this scaling.

In particular, the amplitude for scalar scattering, $M(SSSS)$, is the worst-behaved at high energies and violates unitarity at scales of order $\Lambda_5$.  We find that
\be 
M(SSSS) = -\frac{5(1-6c_1-4c_2)^2}{432 \Lambda_5^{10}} stu(s^2+t^2+u^2) + \cdots,
\ee
in agreement with \Ref{Russian}, which calculated this amplitude including just the Fierz-Pauli term.  By choosing $1-6c_1 -4c_2=0$, we can raise the cutoff from $\Lambda_5$ to $\Lambda_4$, so 
\be
M(SSSS) = \frac{3-16d_1-32d_3}{144 \Lambda_4^8} (s^2+t^2+u^2)^2 + \cdots.
\ee
By choosing $3-16d_1-32d_3=0$, we can then further raise the cutoff from $\Lambda_4$ to $\Lambda_3$.     Notably, these choices of parameters are consistent with \Eq{eq:params}, which we expected due to the improved cutoff in ghost-free massive gravity.  This agreement is a non-trivial check that our calculation of the scattering amplitudes is correct.  

Plugging in all the parameters of ghost-free massive gravity from \Eq{eq:params}, we find improved high-energy behavior scaling as
\be
\begin{aligned}
&M(TTTT) \sim s,&  &M(TVTV) \sim s^2,& &M(TSTS) \sim s^2 , \\
&M(VVVV) \sim s^3,&  &M(VSVS) \sim s^3,& &M(SSSS) \sim s^3 .
\end{aligned}
\ee
From our explicit amplitudes, we find that there is no possible combination of parameters in the action \eqref{eq:Vdef} whereby the high-energy scaling of all amplitudes is $s^2$; if such a combination existed, it would raise the cutoff further.  In particular, $M(VSVS)$ always scales as  $\sim s^3$ or worse.   This agrees with \Ref{Schwartz:2003vj}, which argued that high-energy scaling of $\sim s^2$ is impossible. 

After plugging in \Eq{eq:params}, the leading behavior of the all-scalar amplitude is
\be 
M(SSSS) = -\frac{1-4c_3+ 36c_3^2+64 d_5}{6 \Lambda_3^6}stu + \cdots,
\label{eq:Agal}
\ee
which vanishes for $(c_3,d_5) = (1/6, -1/48)$,  a parameter choice that indeed results in non-interacting scalars in the decoupling limit of the $\Lambda_3$ theory \cite{dRG}.    As a highly non-trivial consistency check, we have verified that the leading high-energy behavior of $M(SSSS)$ in \Eq{eq:Agal} is equal to the scattering amplitude for pure Galileons---including signs and numerical factors---as is mandated by the Goldstone equivalence theorem.

For the remainder of this paper, we assume the parameter choice in \Eq{eq:params}, corresponding to ghost-free massive gravity.

\section{Derivation of Constraints}

\label{sec:constraints}

In this section, we briefly review the mechanics of analytic dispersion relations for amplitudes and their relation to positivity.  We then present our results constraining the parameter space of massive gravity.

\subsection{Analytic Dispersion Relations}

For our analysis, we apply analytic dispersion relations to the amplitude $M(s,t)$, for now dropping the labels for the external polarizations.  As noted previously, the forward amplitude $M(s,0)$ is well-defined since $t$-channel singularities are regulated by the graviton mass $m$.   To begin, consider the contour integral
\be 
f = \frac{1}{2\pi i}\oint_\Gamma {\rm d}s\; \frac{M(s,0)}{(s-\mu^2)^{3}},
\label{eq:fdef}
\ee
where $\mu^2$ corresponds to an arbitrary mass scale chosen in the interval $0 < \mu^2 < 4m^2$. The reason for this stipulation will become clear shortly. 

\begin{figure}[t]
\begin{center}
\includegraphics[height=0.45\textwidth]{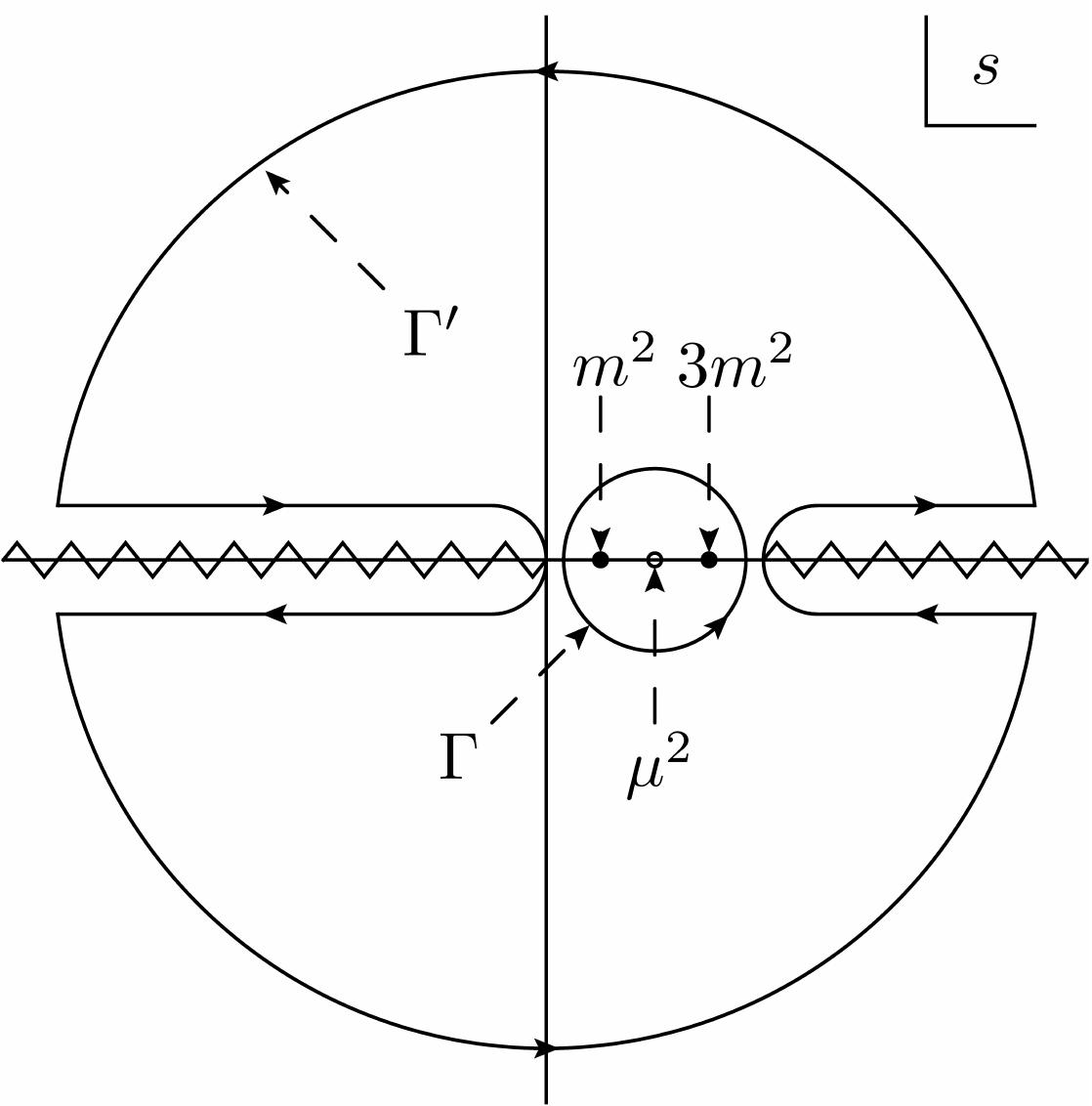}
\end{center}
\vspace{-5mm}
\caption{Diagram of the analytic structure of the forward amplitude in the complex $s$ plane. The simple poles at $s=m^2$ and $3m^2$ and the branch cuts starting at $s=4m^2$ and $0$ correspond to resonances and multi-particle thresholds in the $s$- and $u$-channels, respectively. The scale $\mu^2$ in the dispersion relation is chosen here to be at the symmetric point $\mu^2=2m^2$. The contours $\Gamma$ and $\Gamma'$ referred to in \Eqs{eq:fdef}{eq:fdef2} are also depicted.}
\label{fig:contour}
\end{figure}

 At tree-level, $M(s,t)$ has singularities from massive graviton exchange at $s,t,u=m^2$, which in the forward limit generate simple poles at $s=m^2$ and $ s=3m^2$. Beyond tree-level, branch cuts arise from multi-particle production, which in the forward limit run from $s=4m^2$ to $+\infty$ and from $s=0$ to $-\infty$. The contour $\Gamma$ in \Eq{eq:fdef} is chosen to be a circle of radius at least $m^2$ and at most $2m^2$, centered on $s=2m^2$, so that the contour contains the points $s=m^2$, $s=3m^2$, and $s=\mu^2$, as depicted in \Fig{fig:contour}.

We now use Cauchy's theorem to deform the contour $\Gamma$ into a new contour $\Gamma'$ shown in \Fig{fig:contour}, which runs just above and below the real $s$ axis for $s<0$ and $s>4m^2$, plus a boundary contour at infinity. Assuming the Froissart unitarity bound \cite{Froissart,Martin}, the forward amplitude grows sufficiently slowly with $s$ that the boundary contribution at infinity vanishes \cite{IRUV,us}. Thus,
\be
f = \frac{1}{2\pi i} \oint_{\Gamma'} {\rm d}s\; \frac{M(s,0)}{(s-\mu^2)^{3}} = \frac{1}{2\pi i}\left( \int_{-\infty}^0 + \int_{4m^2}^\infty\right){\rm d}s \frac{{\rm Disc}\,M(s,0)}{(s-\mu^2)^{3}},
\label{eq:fdef2}
\ee
where ${\rm Disc}\,M(s,0)=M(s+i\epsilon,0)-M(s-i\epsilon,0)$ for real $s$ and infinitesimal positive $\epsilon$. For the integral over the negative real $s$ axis, we switch variables to $u=4m^2 - s$, yielding
\be
\begin{aligned}
f &= \frac{1}{2\pi i}\int_{4m^2}^\infty {\rm d}u \frac{{\rm Disc}\,M(4m^2-u,0)}{( 4m^2-u-\mu^2)^3} + \frac{1}{2\pi i}\int_{4m^2}^\infty {\rm d}s \frac{{\rm Disc}\,M(s,0)}{(s-\mu^2)^3}\\
&=\frac{1}{2\pi i} \int_{4m^2}^{\infty} {\rm d}s \left[ \frac{1}{(s-\mu^2)^3} + \frac{1}{(s+\mu^2 - 4m^2)^3}\right]{\rm Disc}\,M(s,0)\\
&=\frac{1}{\pi} \int_{4m^2}^{\infty} {\rm d}s \left[ \frac{1}{(s-\mu^2)^3} + \frac{1}{(s+\mu^2 - 4m^2)^3}\right]{\rm Im}\,M(s,0).\label{eq:ftoImA}
\end{aligned}
\ee
In the second line, we applied the definition ${\rm Disc} \, M(4m^2 - u,0) = M(4m^2 - u + i\epsilon,0) - M(4m^2 - u - i\epsilon,0)$, followed by crossing symmetry,
$M(u,0) = M(4m^2 - u,0)$, thus yielding ${\rm Disc}\,M(4m^2 - u) = M(u - i\epsilon) -M(u + i\epsilon) = -{\rm Disc}\,M(u)$, and then relabeled $u$ to $s$ as a dummy variable.  In the third line, we used the Schwarz reflection principle $M(s^*,0)=[M(s,0)]^*$, so for real $s$ we have ${\rm Disc}\,M(s,0) = 2i\,{\rm Im}\,M(s,0)$. Finally, by applying the optical theorem, $ {\rm Im} \, M(s,0)=s\sigma(s)\sqrt{1-4m^2/s}$, we obtain our final expression,
\be 
f  =\frac{1}{\pi} \int_{4m^2}^{\infty} {\rm d}s \,\sigma(s)\left[ \frac{s}{(s-\mu^2)^3} + \frac{s}{(s+\mu^2 - 4m^2)^3}\right]\sqrt{1-\frac{4m^2}{s}} >0,
\label{eq:fgreat}
\ee
where for an interacting theory the total cross-section $\sigma(s)$ is strictly positive. Since the integration region is restricted to $s>4m^2$ and we stipulated earlier that $0 < \mu^2 <4m^2$, the expressions in brackets and under the radical are strictly positive so $f$ is as well. 

We have applied well-known analytic dispersion relations to prove that $f>0$.  Crucially, from \Eq{eq:fdef} we can derive $f$ purely from the low-energy effective theory, so
\be 
f = \left( \underset{\;\;s=m^2}{\rm Res} \left[\frac{M(s,0)}{(s-\mu^2)^{3}}\right] +  \underset{\;\;s=3m^2}{\rm Res} \left[\frac{M(s,0)}{(s-\mu^2)^{3}}\right] + \underset{\;\;s=\mu^2}{\rm Res} \left[\frac{M(s,0)}{(s-\mu^2)^{3}}\right] \right)_{\rm EFT} >0,
\label{eq:fIR}
\ee
where for emphasis we have included a subscript indicating that all quantities should be computed within the low-energy effective theory, {\it not} the full theory.  There is, however, a shortcut to this calculation: since the poles of the low-energy scattering amplitude are known, we know by Cauchy's theorem that \Eq{eq:fIR} can be calculated in a single step by computing the negative of its residue at large $s$,
\be 
f = -\left( \underset{\;s=\infty}{\rm Res} \left[\frac{M(s,0)}{(s-\mu^2)^{3}}\right] \right)_{\rm EFT} >0,
\label{eq:fIR2}
\ee
which is our final expression for $f$. 

Conveniently, we can show that $f$ is $\mu^2$-independent for ghost-free massive gravity.  In particular, we saw earlier that fixed-angle scattering in ghost-free massive gravity scales as $s^3$.  The only crossing-symmetric invariant at this order, $stu$, vanishes in the forward limit, so forward scattering scales as $s^2$.    At large $s$ we can expand $1/(s-\mu^2) = 1/s + {\cal O}(\mu^2/s^2)$, in which case only the $\mu^2$-independent piece of \Eq{eq:fIR2} contributes. 
 We have verified this to be the case in our explicit amplitudes.

Now we can reintroduce the dependence on the external polarization data.  Since the general amplitude is a quartic form in the polarizations $(\alpha, \beta, \gamma, \delta)$, the forward, crossing-symmetric amplitude is a real quartic form in $(\alpha, \beta)$.  As $f$ is a residue of the latter, it takes the form
\be 
f(\alpha,\beta) = \sum_{ijkl} f(ijkl) \alpha_i \beta_j  \alpha_k \beta_l >0.
\ee 
Obviously, $f(ijkl)$ is symmetric under $i \leftrightarrow k$ and $j \leftrightarrow l$ due to the structure of the quartic form and also under $ik \leftrightarrow jl$ from exchange of the two incoming particles; that is,
\be 
f(ijkl) = f(kjil) = f(ilkj) = f(jilk).\label{eq:symmetries}
\ee 
In principle, these symmetries leave $f(ijkl)$ with 120 independent components, but as we will see, many of these are zero for the physical amplitude.

In the next subsection, we present $f(ijkl)$ and map the positivity bound from analytic dispersion relations onto the parameter space of massive gravity. We begin by studying ``definite-helicity'' gravitons described by pure tensor, vector, or scalar polarizations.  Afterwards, we consider the ``indefinite-helicity'' case in which we are scattering superpositions of these states.

\subsection{Bounds from Definite-Helicity Scattering}

To begin, we consider the scattering of definite-helicity gravitons, corresponding to external polarizations that are purely tensor, vector, or scalar.  
Remarkably, for most combinations of definite-helicity modes, we find that the relative angles between polarizations drop out of our expressions. Writing
\be  
\begin{aligned}
\\
f(1111)= f(1212) = f(2222) &= f(TTTT)\\
f(1313) = f(1414) = f(2323) = f(2424) &= f(TVTV)\\
f(1515) = f(2525) &= f(TSTS)\\
f(3333) = f(4444) &= f(VVVV)_+\\
f(3434) &= f(VVVV)_-\\
f(3535) = f(4545) &= f(VSVS)\\
f(5555) &= f(SSSS),\\
\end{aligned}
\label{eq:polarizedf}
\vspace{5mm}
\ee
expressed in terms of $f$ for various scattering combinations of $T$, $V$, and $S$, we find, via explicit calculation, that
\be 
\begin{aligned}
\\
f(TTTT) =&\; \frac{1}{\Lambda_2^4}  \\
f(TVTV) =&\; \frac{5-12c_3}{4\Lambda_2^4}   \\
f(TSTS) =&\; \frac{5-12c_3}{3\Lambda_2^4} \\
f(VVVV)_+ =&\; \frac{5 + 72 c_3 -240c_3^2}{16\Lambda_2^4} \\
f(VVVV)_- =&\; \frac{23-72c_3+144 c_3^2+192d_5}{16\Lambda_2^4} \\
f(VSVS) =&\; \frac{91 - 312 c_3 + 432 c_3^2 + 384 d_5}{48\Lambda_2^4}  \\
f(SSSS) =&\; \frac{14 - 12 c_3 - 36 c_3^2 + 96 d_5}{9\Lambda_2^4} .
\label{eq:answer}
\end{aligned}
\vspace{5mm}
\ee
Note that only in the case of all-vector scattering does $f$ depend on the relative angle between external polarizations.  For this reason, we had to define both $f(VVVV)_+$ and $f(VVVV)_-$, corresponding vector polarizations that are parallel and orthogonal, respectively.  In contrast, the all-tensor case $f(TTTT)$, for example, is independent of the relative angle between the incoming tensor polarizations.

\begin{figure}[t]
\begin{center}
\hspace*{2cm}\includegraphics[height=0.5\textwidth]{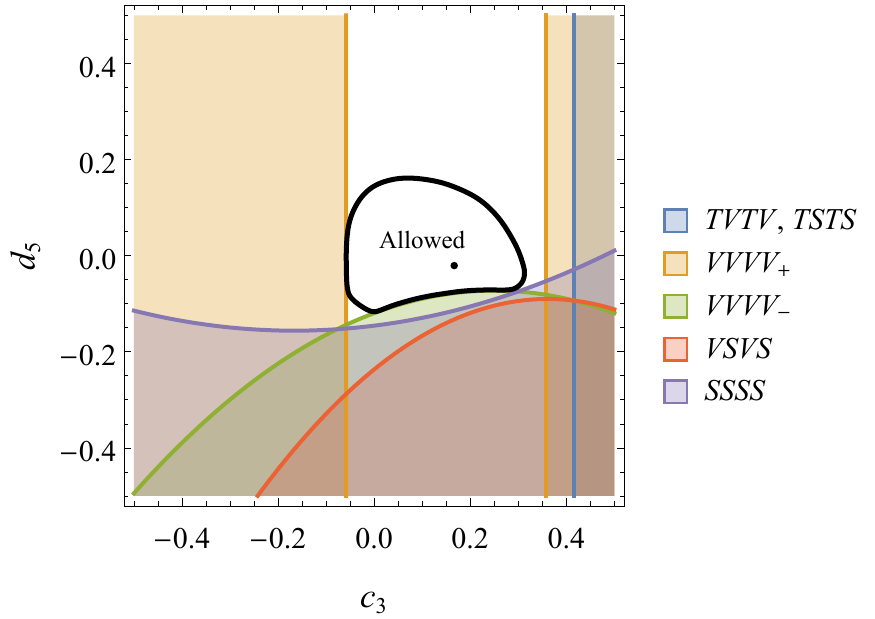}
\end{center}
\vspace{-5mm}
\caption{Regions in the $(c_3,d_5)$ parameter space of ghost-free massive gravity excluded by analyticity bounds on scattering of definite-helicity gravitons.  The tensor, vector, and scalar modes are denoted by $T$, $V$, and $S$, respectively, and the $\pm$ delineation indicates vector polarizations that are parallel or orthogonal, respectively.
Ultimately, by considering indefinite-helicity scattering, we will further restrict the allowed region of parameter space to that within the black curve. 
The dot marks the parameter choice $(c_3, d_5)=(1/6,-1/48)$, which corresponds to a free scalar sector in the decoupling limit.\vspace{1cm}}
\label{fig:exclusions}
\end{figure}

\pagebreak
To obtain new positivity bounds, we simply demand that $f>0$ for all polarization combinations in \Eq{eq:answer}.  These constraints can be cast as an excluded region in $(c_3, d_5)$ space, as shown in \Fig{fig:exclusions}. As one can see, considering the scattering of modes that are pure tensor, vector, or scalar is alone enough to rule out much of the parameter space of massive gravity, except for a strip in $d_5$ for certain values of $c_3$. In order to obtain the most stringent possible bounds, we turn to the question of scattering indefinite-helicity states in the next subsection, which will restrict the allowed parameter space to the region inside the black curve in \Fig{fig:exclusions}.

\subsection{Bounds from Indefinite-Helicity Scattering}

In general, it is possible to scatter arbitrary superpositions of tensor, vector, and scalar modes, corresponding to generic real unit vectors $\alpha$ and $\beta$.  Our calculation shows that all $f(ijkl)$ vanish except for those in \Eq{eq:polarizedf}, together with
\be\begin{aligned}
f(1133) = f(1144) = f(2233) = f(2244) &= -\frac{3(1-4c_3)^2}{8\Lambda_2^4} \\
f(1155) = f(2255) &= \frac{-1+8c_3 - 24c_3^2 - 16d_5}{2\Lambda_2^4}\\
f(1335) = -f(1445) = f(2345) = f(2435)  &= \frac{\sqrt{3}(1-12c_3)^2}{96\Lambda_2^4}\\
f(1353) = -f(1454) = f(2354) &= \frac{\sqrt{3}(1 - 8c_3 + 48c_3^2 + 64d_5)}{16\Lambda_2^4}\\
f(3344) &= \frac{-9 + 72c_3 - 192c_3^2 - 96d_5}{16\Lambda_2^4}\\
f(3355) = f(4455) &= \frac{-17 + 136c_3 - 336c_3^2}{32\Lambda_2^4},\\
\end{aligned}\label{eq:quarticform2}
\ee
along with the $f(ijkl)$ related to these by the symmetries in \Eq{eq:symmetries}.
Varying $(\alpha,\beta)$ corresponds to different scattering experiments in which the scattered particles are various superpositions of polarizations. Imposing analyticity constraints on the amplitude for all possible scattering processes---that is, marginalizing over all possible choices of $(\alpha,\beta)$---implies positivity bounds on the massive graviton parameter space that are much stronger than the bounds derived in the previous subsection.  

For example, consider gravitons that are maximal superpositions of scalar and tensor,
\be
\alpha_i = \beta_i = \frac{1}{\sqrt{2}}(\cos{\phi}, \sin{\phi},0,0,1).
\ee
For any value of $\phi$, the corresponding scattering amplitude yields
\be
f(\alpha,\beta) =  \frac{35 + 60 c_3 - 468 c_3^2 - 192 d_5}{36\Lambda_2^4}.
\ee
Requiring positivity of $f$ then excludes arbitrarily large values of $d_5$, irrespective of $c_3$.  In terms of the $(c_3,d_5)$ parameter space, this example bound already eliminates all but a compact region of the semi-infinite strip of the parameter space permitted by the definite-helicity graviton scattering bounds shown in \Fig{fig:exclusions}.

To place the most stringent bounds from analytic dispersion relations, we must find all points in $(c_3, d_5)$ for which $f$ is positive for all $(\alpha,\beta)$. That is, we must marginalize over all  choices of external polarizations. Unfortunately, there is no analytic prescription for determining the positivity of quartic forms.  While this algebraic problem is strongly NP-hard \cite{NPhard}, it can be recast as a dynamical problem \cite{math} that is numerically tractable.  In particular, let us repackage $(\alpha,\beta)$ into a new ten-dimensional ``coordinate'', 
\be 
X_I = (\alpha_1,\alpha_2,\alpha_3,\alpha_4,\alpha_5,\beta_1,\beta_2,\beta_3,\beta_4,\beta_5),
\ee
relaxing the normalization constraint $\alpha^2 = \beta^2 =1$.  Next, we assume that $X_I$ evolves in time $t$ according to an equation of motion,
\be
\frac{{\rm d}{X}_I}{{\rm d}t} = - \frac{\partial f}{\partial X_I} . \label{eq:dynamical}
\ee
This immediately implies that
\bea
\frac{{\rm d}f}{{\rm d}t} = - \sum_I \frac{\partial f}{\partial X_I}\frac{\partial f}{\partial X_I} \leq 0,
\eea
so $f$ is non-increasing over time. Meanwhile, we know that as long as $X_I \neq0$, then $\partial f/\partial X_I\neq 0$ since $f$ is quartic in $X_I$.  It thus follows that ${\rm d}f/{\rm d}t < 0$ strictly everywhere away from $X_I = 0$, {\it i.e.}, $f$ will decrease monotonically at all $X_I$ except the origin.  If there is a direction in which $f$ is unbounded from below, then time evolution will drive it arbitrarily negative.  On the other hand, a positive definite $f$ will of course remain positive forever.    As a result, $f$ is positive definite if and only if $f$ is stable under the time evolution of $X_I$.  

\begin{figure}[t]
\begin{center}
\includegraphics[height=0.5\textwidth]{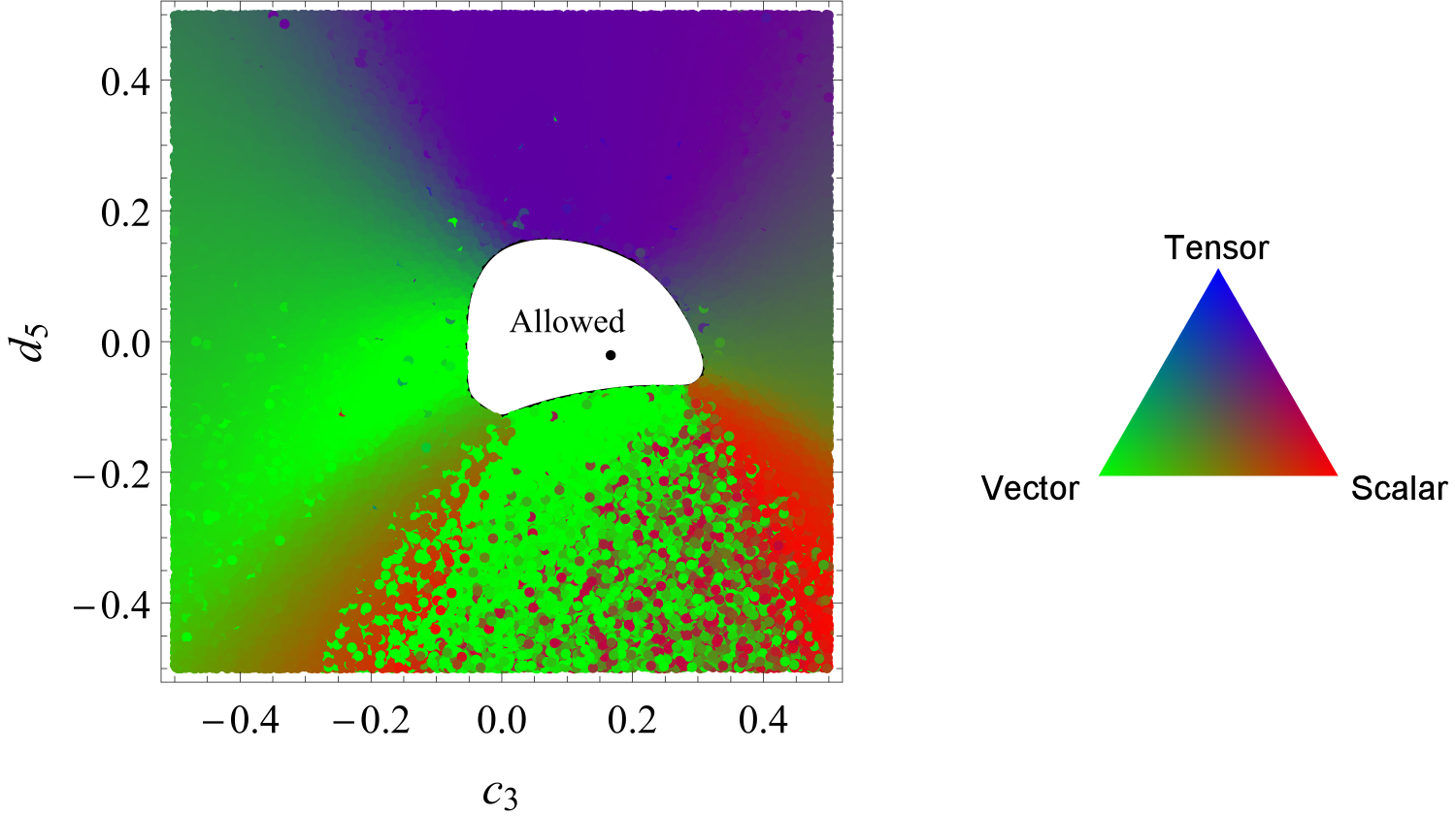}
\end{center}
\vspace{-5mm}
\caption{Region of $(c_3,d_5)$ parameter space for ghost-free massive gravity excluded by analyticity bounds on scattering of indefinite-helicity gravitons.   
Each colored point corresponds to a theory excluded by a scattering process that violates analytic dispersion relations.  As noted in text, such violations can be diagnosed by evolving a particular dynamical system that tends toward scattering processes of gravitons of similar polarization.  The specific color---plotted in blue, green, and red---corresponds to the power of each polarization in tensors ($\alpha_1^2 + \alpha_2^2$ and $\beta_1^2+\beta_2^2$), vectors ($\alpha_3^2+\alpha_4^2$ and $\beta_3^2 + \beta_4^2$), and scalars ($\alpha_5^2$ and $\beta_5^2$).
The allowed region is shown in white and the black dot marks the choice that corresponds to a free Galileon.
}
\label{fig:dynamical}
\end{figure}

Concretely, for a given numerical choice of $(c_3,d_5)$, we initialize a random value of  $X_I(t_{\rm init})$, evolve in time to $X_I(t_{\rm final})$, and then check whether $f(t_{\rm final})$ is negative.  If so, then the polarization choices given by $X_I(t_{\rm final})$, suitably normalized, contradict the analyticity argument.  Thus the parameter point $(c_3,d_5)$ is inconsistent and we discard it.  If $f(t_{\rm final}) \geq 0$, the parameter point remains a possible viable theory.
Iterating many times, we are able determine a definitive region in $(c_3, d_5)$ that is excluded by analytic dispersion relations for all possible graviton scattering configurations.  

The result of this calculation is that $(c_3,d_5)$ are confined to a small compact region, as shown in \Fig{fig:dynamical}.  Here each colored point corresponds to a point in parameter space for which our algorithm has determined a violation of analytic dispersion relations.  The color of the point encodes the power distribution in the tensor, vector, and scalar components of the corresponding polarization excluding the point. Interestingly, we find that for many of the points that violate positivity, the numerical algorithm tends to converge to scattering processes in which the two scattered gravitons have the same power distribution.

\section{Implications for Massive Gravity}

\label{sec:galileon}

Our bounds exclude most of the parameter space for ghost-free massive gravity, subject to the assumptions of analyticity and unitarity of the theory.  While this is in part a negative result,  the existence of a finite allowed region is actually encouraging, especially given the checkered history of the scalar mode of massive gravity---the so-called Galileon.  

As demonstrated early on, the Galileon is a remarkable effective theory in and of itself  \cite{Rattazzi}.  The model is uniquely fixed by an extended shift symmetry that highly constrains allowed interactions, limiting the action to a set of five Galilean-invariant operators in four dimensions. The Galileon is by construction ghost-free, which is natural since it describes the scalar mode of ghost-free massive gravity.  Moreover it supports interesting cosmological solutions \cite{GalileonCosmology,Acceleration,dRHCosmology} and has scattering amplitudes with unique infrared properties \cite{Soft}.

On the other hand, it has long been known that the Galileon actually violates positivity bounds derived from analytic dispersion relations \cite{Rattazzi,Positivity,IRUV}.  The reason is simple: the extended shift symmetry of the Galileon simply forbids interactions of the form $(\partial S)^4$, which induce $s^2$ contributions to the amplitude.  Galileon interactions are instead of the form $(\partial S)^2 (\partial\partial S)^2$, which mandates strict $s^3$ behavior of the fixed-angle amplitude, with no subleading corrections.  In turn, the only crossing-symmetric invariant of this type is $stu$, which is zero in the forward limit.  Consequently, $f(SSSS)=0$, which is not strictly positive, contradicting \Eq{eq:fgreat}.\footnote{Note that in certain conformal variations of the Galileon \cite{Positivity}, the theory is modified, permitting $(\partial S)^4$ corrections that allow accordance with analyticity constraints.}  Thus, the pure Galileon theory is marginally excluded by analyticity bounds.

These results are consistent with our own because the Galileon only describes the scalar mode of massive gravity in the limit of Goldstone equivalence.  In contrast, our results automatically incorporate all contributions coming from the tensor and vector modes as well. More importantly, our calculation implicitly includes subleading corrections to Goldstone equivalence that scale as higher powers in $m^2 /s$ relative to the pure Galileon result.  Thus, while the leading behavior of \Eq{eq:Agal} scales as $stu$ as expected, there are subleading corrections at order $s^2$ that are nonzero.  Since the pure Galileon is only marginally inconsistent with analyticity bounds, the right choice of $(c_3, d_5)$ can tip the scales.  In this sense, our calculation shows explicitly that the pathologies of the Galileon are remedied when embedded in a full theory of massive gravity.

\section{Conclusions}
\label{sec:conclusions}
\vspace{-2.1mm}

In this paper, we have used the principles of unitarity and analyticity of scattering amplitudes to bound the general effective theory of a massive graviton.  We have shown that the consistency of massive graviton scattering significantly constrains the parameter space of ghost-free massive gravity. Analyticity bounds have been analyzed in other contexts, both in non-gravitational \cite{IRUV,OldAnalyticity} and more recently gravitational \cite{IRUV,IRWGC,us} theories. Such analyses provide useful criteria for charting the boundary between the landscape and the swampland.  As the principles from which these bounds are derived are infrared properties, they apply to any well-behaved ultraviolet completion obeying the canonical axioms of field theory, irrespective of what the ultimate theory of quantum gravity may be.
\vspace{-1mm}
\begin{center} 
 {\bf Acknowledgments}
 \end{center}
 \vspace{-1mm}
 \noindent 
We thank Brando Bellazzini, Kurt Hinterbichler, and Rachel Rosen for useful discussions and comments.
C.C.~is supported by a Sloan Research Fellowship and a DOE Early Career Award under Grant No.~DE-SC0010255. G.N.R.~is supported by a Hertz Graduate Fellowship and a NSF Graduate Research Fellowship under Grant No.~DGE-1144469.

\bibliography{massive_gravity}
\bibliographystyle{utphys}
\end{document}